\documentclass{emulateapj}
\usepackage{apjfonts}

\newcommand{\mum}{\ifmmode{\rm \mu m}\else{$\mu$m}\fi}
\newcommand{\iso}{{\em ISO}}

\begin{document}

\title{The detection of silicate emission from quasars at 10 and 18 microns}

\author{
Lei~Hao\altaffilmark{1},
H.~W.~W.~Spoon\altaffilmark{1,2},
G.~C.~ Sloan\altaffilmark{1},
J.~A.~Marshall\altaffilmark{1},
L.~Armus\altaffilmark{3},
A.~G.~G.~M.~Tielens\altaffilmark{4},
B.~Sargent\altaffilmark{5},
I.~M.~van~Bemmel\altaffilmark{6},
V.~Charmandaris\altaffilmark{1,7,8},
D.~W.~Weedman\altaffilmark{1},
J.~R.~Houck\altaffilmark{1}
\email{haol@isc.astro.cornell.edu}
}
\altaffiltext{1}{Cornell University, Astronomy Department, 
  Ithaca, NY 14853-6801}
\altaffiltext{2}{Spitzer Fellow}
\altaffiltext{3}{Caltech, Spitzer Science Center, MS 220-6, Pasadena, CA 91125}
\altaffiltext{4}{SRON National Institute for Space Research and
  Kapteyn Institute, P.O. Box 800, 9700 AV Groningen, The Netherlands}
\altaffiltext{5}{University of Rochester, Department of Physics and
  Astronomy, Rochester, NY 14627}
\altaffiltext{6}{Space Telescope Science Institute, 3700 San Martin Drive,
  Baltimore, MD 21218}
\altaffiltext{7}{University of Crete, Department of Physics,
  P.~O. Box 2208, GR-71003, Heraklion, Greece}
\altaffiltext{8}{Chercheur Associ\'{e} Obs. de Paris, 61 Ave. de 
  l'Observatoire, F-75014 Paris, France}

\slugcomment{Submitted to Astrophysical Journal Letters, 2005 March 24,
accepted 2005 April 19}

\begin{abstract}

We report the spectroscopic detection of silicate emission at 10 and
18\,\mum\ in five PG quasars, the first detection of these two features in
galaxies outside the Local Group.  This finding is consistent with the
unification model for Active Galactic Nuclei (AGNs), which predicts
that an AGN torus seen pole-on should show a silicate emission feature
in the mid-infrared.  The strengths of the detected silicate emission
features range from 0.12 to 1.25 times the continuum at 10\,\mum\ and
from 0.20 to 0.79 times the continuum at 18\,\mum.  The silicate grain
temperatures inferred from the ratio of 18-to-10\,$\mu$m silicate
features under the assumption of optically thin emission range from
140 to 220\,K.

\end{abstract}

\keywords{galaxies: active --- quasars: emission lines --- galaxies: ISM ---
infrared: galaxies} 

%%%%%%%%%%%%%%%%%%%%%%%%%%%%%%%%%%%%%%%%%%%%%%%%%%%%%%%%%%%%%%%%%%%%%%%%%%

\section{Introduction}

Active galactic nuclei (AGNs) are broadly classified in two 
types.  Type 1 AGNs display broad hydrogen emission lines in 
the optical, while Type 2 AGNs do not.  The AGN unification 
model \citep[e.g.][]{ant93} ties these two types together.  An 
optically and geometrically thick dusty torus surrounds a 
central black hole, accretion disk and broad-line emission 
region.  Sources viewed face-on are recognized as Type 1, 
while edge-on sources are Type 2.  Observations at many
wavelengths support this scenario \citep[e.g.][]{am85}, but 
until now there was a lack of constraining data in the
mid-infrared.

Silicate dust is a major component of the interstellar medium 
in the Milky Way and other galaxies.  It produces two spectral 
features in the infrared, a ``10\,\mum\ feature'' which arises 
from a Si$-$O stretching mode, and an ``18\,\mum\ feature'' from 
an O$-$Si$-$O bending mode \citep[e.g.][]{kt73}.  Emission or 
absorption from silicate dust is a dominant feature of the 
infrared spectrum produced by most mass-losing evolved stars, 
interstellar clouds, H{\sc ii} regions, and circumstellar disks 
around young stellar objects (YSOs).  Silicate absorption is 
also detected in dusty starburst galaxies \citep{rig99}, 
ultraluminous infrared galaxies \citep[ULIRGs;][]{gen98, 
tra01, spo04, arm04}, and a sample of mid-infrared detected, 
optically invisible, high-luminosity galaxies with redshifts 
of 1.7 $<$ z $<$ 2.8 \citep{hou05}.

Type 2 AGNs show silicate dust in absorption as expected, but the
mid-infrared spectroscopic data for Type 1 AGNs have been scarce. 
Recently, \cite{sie05} have presented spectra of two
quasars with just the 10\,\mum\ feature in emission.  While radiative 
transfer models clearly predict silicates to be in emission for Type 1 
sources \citep{ld93, pk92, pk93, gd94, err95, nie00, nie01, nie02, 
vbd03, dvb05}, the absence of the silicate emission features at 10 and
18\,\mum\ has led to models without strong silicate emission.
\cite{ld93} and \cite{vbd03} have suggested that larger grains
dominate the grain-size distribution.  \cite{nie02}, instead,
proposed that if the torus is clumpy, the 10\,\mum\ silicate emission
can be sufficiently suppressed, although \cite{dvb05} have challenged
this suggestion.

Here, we report the first spectroscopic detection of both
silicate features at high signal to noise in five PG quasars.
Quasars are the high luminosity counterparts of Seyfert 1 galaxies,
showing the same broad emission lines that define Type 1 AGNs.

%%%%%%%%%%%%%%%%%%%%%%%%%%%%%%%%%%%%%%%%%%%%%%%%%%%%%%%%%%%%%%%%%%%%%%%%%%

\section{Observations and data reduction}

%%%%%%%%%%%%%%%%%%%%%%%%%%%%%%%%%%%%%%%%%%%%%%%%%%%%%%%%%%%%%%%%%%%%%%%%%%

The five sources presented in this letter (PG 0804+761, PG 1211+143,
PG 1351+640, I Zw 1 [=PG 0050+124] and 3C~273 [=PG 1226+023]) were
selected from a sample of 12 AGNs based
upon their prominant silicate emission features at both 10 and
18\,\mum. The observations were part of the Guaranteed Time
Observation (GTO) program of the Infrared Spectrograph
(IRS)\footnote{The IRS was a collaborative venture between Cornell
University and Ball Aerospace Corporation funded by NASA through the
Jet Propulsion Laboratory and the Ames Research Center}\citep{hou04}
on the Spitzer Space Telescope \citep{wer04}.  Table \ref{tab1} lists
the basic properties of these targets, along with their observation
dates and on-source integration times.  All of them have broad
H$\beta$ lines \citep{bg92} and therefore are classified as Type 1
AGNs. Two of them, 3C~273 and PG 1211 are radio-loud quasars and the
rest are radio-quiet \citep{san89}.  All five quasars are very
luminous, as can be seen from the bolometric luminosities in
Table\,\ref{tab1} \citep[from][]{san89}.

\begin{table} % Table 1
\caption{Properties of Sources\label{tab1}}
\begin{tabular}{lccccc}
\colrule
\colrule
 & PG 1351 & I Zw 1 & 3C 273 & PG 1211 & PG 0804 \\
\colrule
AOR key                              & 3760640   & 3761920  & 4978176  & 3760896  & 9074944 \\
Date observed                        & 2004/4/17 & 2004/1/7 & 2004/1/6 & 2004/1/7 & 2004/3/1\\
Integration time (s)                 & 408       & 224      & 336      & 408      & 408\\
Redshift                             & 0.0881    & 0.0611   & 0.158    & 0.0808   & 0.0999 \\
Lum. distance (Mpc)\tablenotemark{a} & 398       & 271      & 749      & 363      & 455 \\ 
log($L_{\rm bol}/L_{\odot}$)         & 12.14     & 12.20    & 13.44    & 12.44    & 12.56\\
F$_{15\,\mum}$ (Jy)\tablenotemark{b} & 0.21      & 0.62     & 0.36     & 0.21     & 0.13\\
\colrule
\end{tabular}
\tablenotetext{a}{assuming H$_0$=71\,km\,s$^{-1}$\,Mpc$^{-1}$, 
$\Omega_M$=0.27, $\Omega_{\Lambda}$=0.73, $\Omega_K$=0}
\tablenotetext{b}{rest frame}
\end{table}

The observations were made with the Short-Low (SL) and Long-Low (LL)
modules of the {\em IRS}.  The spectra were extracted from the 
flatfielded images
provided by the Spitzer Science Center (pipeline version S11.0.2). The
images were background-subtracted by differencing the two SL apertures
and for LL, by differencing the two nod positions.  Spectra were then
extracted and calibrated using the {\em IRS} standard star HR 6348 for
SL and the stars HR 6348, HD 166780, and HD 173511 for LL
\citep{slo05}.

\begin{figure} % Fig. 1
\begin{center}
\resizebox{\hsize}{!}{\includegraphics{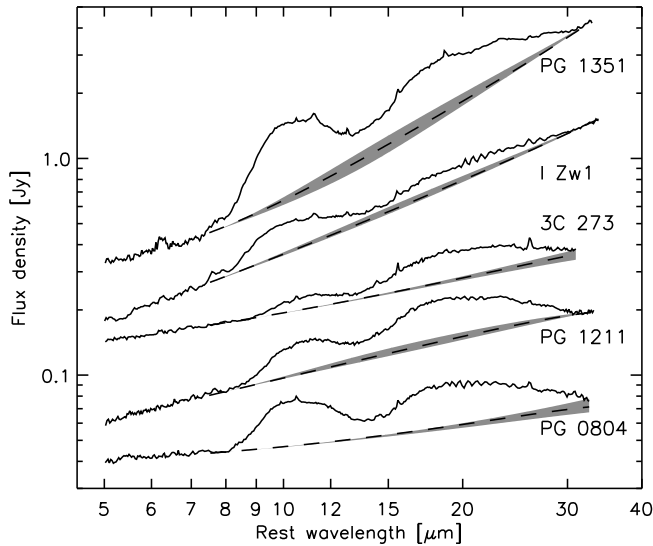}}
\caption{The 5--35\,\mum\ IRS low resolution spectra of five PG
quasars showing silicates in emission.  The pixel-to-pixel 
uncertainty in the data is less than 3\%.  Fringing effects with
amplitudes of up to 5\% are visible in LL order 1 (see text). For 
each spectrum, three choices for the silicate-graphite continuum 
are indicated by the dashed line and grey shaded area. 
The spectra have been scaled for plotting purposes, with scaling
factors of:  8.0 for PG\,1351, 1.0 for I\,Zw1, 0.77 for 3C\,273 
and PG\,1211 and 0.50 for PG\,0804.
\label{fig1}}
\end{center}
\end{figure}

After extraction the orders were stitched, requiring order-to-order
scaling adjustments of less than 5\%.  Finally, the stitched spectra
were scaled to match the observed {\em IRAS} fluxes at 12 and
25\,\mum\ or {\em IRS} blue peak-up flux at 16~\mum.  For 3C 273, we
did not attempt any scaling, since the source varies by up to 0.4
magnitudes at 10.6\,\mum\ on time scales of up to several years
\citep{neu99}.  The scaling factors we applied average 6\%, with
the largest scaling factor of 16\%. The observed sources are point
sources for both IRAS and IRS, therefore scaling to IRAS flux is
justified.  Figure \ref{fig1} presents the calibrated spectra.  Table
\ref{tab1} lists the 15\,\mum\ rest frame fluxes.

\begin{figure} % Fig. 2
\begin{center}
\resizebox{\hsize}{!}{\includegraphics{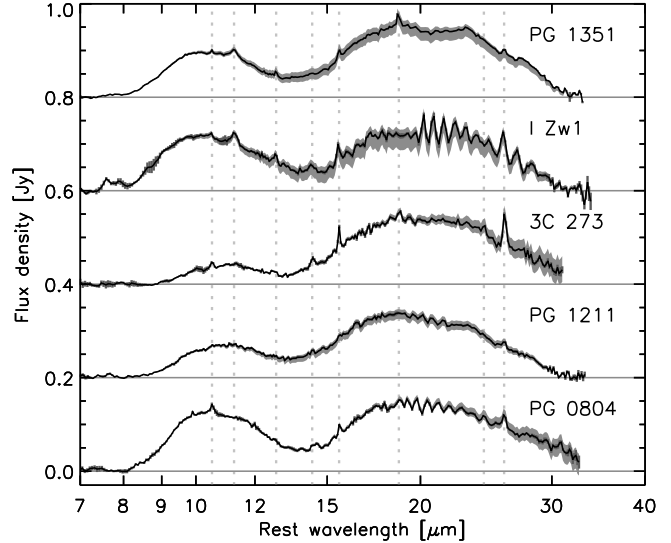}}
\caption{Continuum-subtracted silicate emission spectra of the 
five PG quasars. The spectra have been normalized and offset 
from each other for plotting purposes.  Fringing effects are 
visible in LL order 1 (see text).  The shaded areas indicate the 
uncertainty in the silicate emission spectra as introduced by the 
uncertainty in the observed spectrum and choice of continuum.
Vertical dotted lines indicate the rest wavelengths of emission
features discussed in the text.
\label{fig2}}
\end{center}
\end{figure}

\begin{figure} % Fig. 3
\begin{center}
\resizebox{\hsize}{!}{\includegraphics{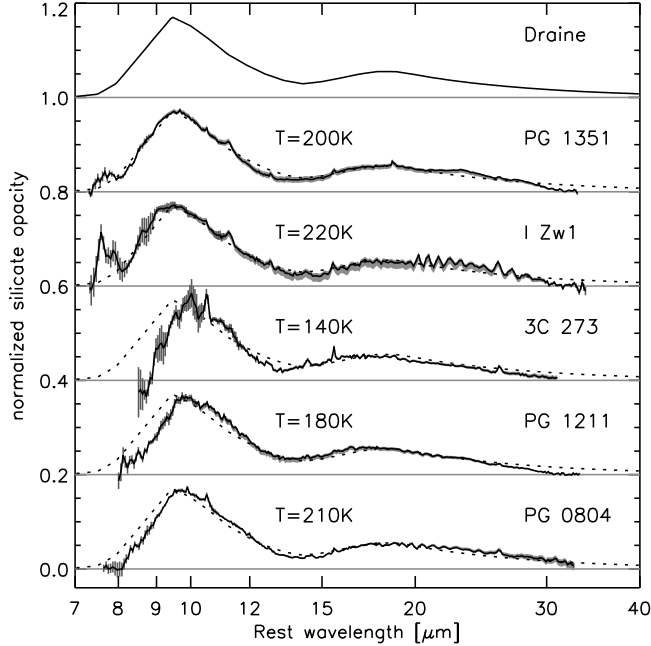}}
\caption{
The opacity spectra for the five PG quasars compared to the 
scaled synthetic silicate opacity curve of \cite{wein01} and 
\cite{li01}.  The opacity spectra from the quasars are derived 
using the adopted silicate grain temperature and scaled and offset 
for plotting purposes.  The shaded areas indicate the uncertainty 
in the silicate profile as introduced by the uncertainty in the 
observed spectrum and choice of continuum.
\label{fig3}}
\end{center}
\end{figure}

In some spectra, most notably in IZw1, residual fringing, which
are due to minor pointing errors, appears between 21 and 30\,\mum\ (the
1st order of the LL module).  We have not defringed the spectra.
Residual artifacts also appear in some spectra in SL order 1
(7.5--14\,\mum).  These take the form of a roughly sinusoidal
deviation from the actual spectrum and are most obvious in the
spectrum of 3C 273 at 8--10\,\mum, where they have a maximum strength
of $\sim$3\% of the total flux.  Because 3C 273 has the weakest
silicate emission feature in our sample, this artifact is even more
pronounced in figures showing the continuum-subtracted spectra.  We
estimate uncertainties by comparing the spectra in the two nod
positions, and these artifacts generally manifest themselves as a
disagreement between the nod positions.  The uncertainties only become
apparant in Figures \ref{fig2} and \ref{fig3} shortward of 11\,\mum\
(dark shaded area). None of these artifacts change the conclusions
reported below.

%%%%%%%%%%%%%%%%%%%%%%%%%%%%%%%%%%%%%%%%%%%%%%%%%%%%%%%%%%%%%%%%%%%%%

\section{Analysis}

The spectra in Figure \ref{fig1} are dominated by two broad emission
features, which we identify as the Si--O stretching mode and the 
O--Si--O bending mode
in silicate dust grains.  The strength of the features
is best appreciated by comparing the peak fluxes to the underlying
silicate-graphite ``continuum''. For PG 1351, I Zw 1 and PG 1211, we
define the continuum using a spline interpolation between the
5--8\,\mum\ spectrum and a point at the extreme red end of the IRS
spectral range, where a change in slope indicates the end of the
18\,\mum\ silicate feature.  For 3C 273 and PG 0804, the spectral
structure at the red end of the IRS spectral range does not show a
similar change in slope.  This may indicate that dust emission is
still contributing at these wavelengths.  Therefore, we have assumed
that the continuum at the red end runs somewhat below the observed
spectrum.  Figure \ref{fig1} plots the resulting spline-interpolated
continua (dashed lines), while Figure \ref{fig2} shows the resulting
continuum-subtracted spectra.
%The effect of slightly different choices of continua is
%illustrated by the grey shaded areas in both Figures.

Comparison of the silicate emission spectra in Figure \ref{fig2}
reveals that the ratio of the 18 and 10\,\mum\ features varies
significantly within the sample.  It is highest for 3C 273 and lowest
for PG 0804 (see Table \ref{tab2}).  Also the onset and center of the
10\,\mum\ features shift significantly from one source to the next, 3C
273 and PG 1211 seem to appear at longer wavelengths. Interestingly,
these two sources happen to be the only radio loud quasars in our
sample. In order to test whether these differences can be attributed
to our choice of continuum, for each source we compared the silicate
emission spectra for two other choices of underlying silicate-graphite
continuum: one above and one below the adopted continuum, enclosing
the shaded areas in Figure \ref{fig1}.  As illustrated by the shaded
areas in Figure \ref{fig2}, this results in an uncertainty of up to
30\% in the peak flux of the 18\,$\mu$m feature and in the ratio of
the 18 to 10\,\mum\ features. In contrast, the onset and center of the
10\,\mum\ features are not affected by different choices of continuum.

\begin{table} % Table 2
\caption{Measured quantities\label{tab2}}
\begin{tabular}{lccccc}
\colrule
\colrule
 & PG 1351 & I Zw 1 & 3C 273 & PG 1211 & PG 0804 \\
\colrule
10$\mu$m-feature-to-cont. ratio\tablenotemark{a}       & 1.25  & 0.38  & 0.12  & 0.55  & 0.60 \\
18$\mu$m-feature-to-cont. ratio\tablenotemark{a}       & 0.79  & 0.20  & 0.36  & 0.56  & 0.56 \\
18$\mu$m\,/\,10$\mu$m feature ratio\tablenotemark{a}   & 1.75  & 1.31  & 3.34  & 1.89  & 1.08 \\
F(30$\mu$m)/F(5.5$\mu$m)              & 11.0  & 6.86  & 2.53  & 2.99  & 1.95 \\
L$_{\rm sil}$/L$_{\rm bol}$           & 0.087 & 0.040 & 0.009 & 0.023 & 0.021\\
10/18$\mu$m cont. color temp (K)\tablenotemark{b}      & 250   & 270   & 320   & 310   & 340  \\
%Silicate temperature (K)             & 200   & 230   & 150   & 180   & 220  \\
Silicate temperature (K)              & 200   & 220   & 140   & 180   & 210  \\
\colrule
\end{tabular}
\tablenotetext{a}{evaluated at the peak of the feature}
\tablenotetext{b}{the color temperature is computed from the adopted continuum at 10 and 18\,\mum\ }
\end{table}

The presence of the silicate features at 10 and 18\,\mum\
allows us to estimate the mean tempeature of the emitting grains by
performing an unweighted $\chi^2$ fit to the continuum-subtracted
spectrum. This temperature is only physically meaningful under the
assumption that the feature emitting region is optically thin. Using
the continuum-subtracted silicate opacity (see top panel in Figure
\ref{fig3}) derived from \cite{wein01} and \cite{li01}, we force the
product of a blackbody and the silicate feature opacities to fit the
observed continuum-subtracted silicate features. This reduces the
problem to a single free parameter: the grain temperature, and avoids 
concerns related to a possible nonthermal contribution to the continuum.  
The resulting silicate grain temperatures (T$_{\rm sil}$) range between
140\,K for 3C 273 and 220\,K for I Zw 1, see Table \ref{tab2}.  These
temperatures will change by as much as 15\,K depending on the choice
for the underlying silicate-graphite continuum.  Uncertainties in the
fitting process may amount to as much as another 10\,K.

Using the derived silicate grain temperature (T$_{\rm sil}$), 
it is possible to compute the silicate opacity spectrum by 
dividing the observed silicate emission spectrum by the 
blackbody spectrum with temperature T$_{\rm sil}$. 
Figure \ref{fig3} presents the resulting opacity spectra 
for our sample.

The top panel in Figure \ref{fig3} shows the synthetic silicate
opacity spectrum after subtraction of the silicate-graphite continuum
component.  The lower panels show the opacity spectra for the five
quasars overlaid on the synthetic profile.  The quasar profiles
generally show good agreement at wavelengths beyond 10\,\mum.  Below
10\,\mum, only PG 1351 and I Zw 1 match the synthetic profile well.
In the other three spectra, the blue wing of the 10\,\mum\ appears
weak, especially in 3C 273 and PG 1211.  Apart from the artifacts in
the 8--10\,\mum\ region of the spectrum of 3C 273 (indicated by the
large error bars in Figure \ref{fig2}; see \S 2) and the presence of
atomic lines and PAH emission in some spectra, the 10\,\mum\ band has
a smooth appearance, showing no sign of a departure from amorphous
grain structure.  Evidence for some crystalline grains do appear at
longer wavelengths, most notably in the spectrum of PG 1351 at
23\,\mum.

The weakness of the blue wing of the 10\,$\mu$m silicate opacity
profile in three of our sources appears to be significant and cannot
be explained by the uncertainties in the observed spectrum (see the
error bars in Figure \ref{fig3}). Grain size and composition,
geometry, optically thick radiative transfer or a combination of these
factors may explain the observed weakness. If our optically thin
assumption is invalid, then the temperatures we have inferred cannot
be interpreted as real dust temperatures. \cite{jaf04} reported a
similar deficiency in the blue wing of the 10um absorption feature in
NGC1068, which they attribute to a different silicate grain
composition.

A close inspection of the spectra in Figures \ref{fig1} and \ref{fig2}
reveals numerous other features.  Those visible in at least one
spectrum in the sample include the PAH emission features at 6.2, 7.7
and 11.2\,\mum, and emission lines at 10.5\,\mum\ ([S{\sc iv}]),
12.8\,\mum\ ([Ne{\sc ii}]), 14.3\,\mum\ ([Ne{\sc v}]), 15.6\,\mum\
([Ne{\sc iii}]), 18.7\,\mum\ ([S{\sc iii}]), and 25.9\,\mum\ ([O{\sc
iv}]).

%%%%%%%%%%%%%%%%%%%%%%%%%%%%%%%%%%%%%%%%%%%%%%%%%%%%%%%%%%%%%%%%%%%%%

\section{Comparison with Models}

Direct comparisons of our observations to the predictions of published 
AGN radiative transfer models are difficult given the limited resolution 
of model spectra in the 5--40\,\mum\ range of the IRS.  Additionally, 
the model output is not tuned to be compared to the observed
properties presented in Table\,\ref{tab2}. However, we can measure the
feature-to-continuum ratio at 10\,\mum\ from the figures presented by
\cite{ld93, gd94, nie02, dvb05}.  For the optically thin dust models
of \cite{ld93} the values range from 0.1 for models where the
grain-size distribution is dominated by larger grains to 1.5 for
models with a standard grain-size distribution.  In the face-on cases
modeled by \cite{gd94}, the values range from near zero for very thick and
compact configurations to 1.0 for more extended torus models.  These
results are similar to those by \cite{nie02} and \cite{dvb05}, but
their highest 10\,\mum\ feature-to-continuum ratios are around 0.7 and
0.4, respectively.  Given the observed range of 0.12 to 1.25 within
our limited sample (Table \ref{tab2}), the model predictions for the
10\,\mum\ feature-to-continuum ratio are in good agreement. A similar
comparison for the 18\,\mum\ feature-to-continuum ratio could not be
undertaken, since the features are not as apparent as the 10\,\mum\
features in the AGN models.

%%%%%%%%%%%%%%%%%%%%%%%%%%%%%%%%%%%%%%%%%%%%%%%%%%%%%%%%%%%%%%%%%%%%%

\section{Discussion and Conclusion}

We have detected the 10 and 18\,\mum\ silicate emission 
features in five PG quasars --- the first detection of both 
silicate features in emission in galaxies outside the Local 
Group and in AGNs in particular.  Given the existence of 
published mid-infrared photometric and spectroscopic data for 
these quasars, our finding may appear surprising at first.
\cite{roc91} obtained 8--13\,\mum\ spectra of 3C 273 and
I Zw 1, but the low contrast of the silicate feature to the 
strong continnum and the lack of coverage outside the 
N-band window prevented them from identifying silicate emission.
The {\it PHT-S} spectrometer on \iso\ \citep{lem96} obtained
spectra of 3C 273, I Zw 1 and PG 0804 \citep{rig99}, 
but the limited wavelength coverage of {\it PHT-S} (only to 
11.6\,\mum) and the redshifts of the quasars allowed only the 
blue side of the 10~\mum\ feature to be visible.  Because of 
these difficulties, silicate emission had remained unidentified
in these five sources until now.

Our detection of silicate emission in Type 1 AGNs provides support for
the unified AGN model \citep{ant93}.  An AGN torus viewed pole-on
should offer an unobstructed line of sight to the torus dust, the
surface of which is heated to temperatures of several hundred to a
thousand Kelvin by the radiation from the central engine.  Emission
from this hot surface should produce a silicate feature in emission
under most circumstances.

Assuming a simple optically thin geometry, we find that the silicate
grain temperature in our sources ranges from 140 to 220\,K, which is
well below the dust sublimation temperature. Detailed comparison of
the derived opacity profiles for our quasars indicates that our simple
model cannot satisfactorally explain the absence of a blue wing in the
10\,\mum\ feature for three of our quasars. This likely points to a
more complex geometry, involving radiative transfer within a
(partially) optically thick environment -- probably an AGN torus. It
may also point to differences in the grain size or composition.  Our
derived silicate grain temperatures should, therefore, only be used as
a characterization of the observed spectra, not as a means to infer
the actual temperature and geometry of the AGN.

From our limited sample of PG quasars (three in addition to the sample
discussed in this Letter) and a few Seyfert 1 galaxies, it is clear
that 10\,\mum\ silicate emission is not a ubiquitous feature in
mid-infared spectra of Type 1 AGNs.  Only 5 out of 12
continuum-dominated AGNs show this feature prominently. Interestingly,
the quasars which show a 10\,\mum\ silicate emission feature all have
bolometric luminosities higher than 10$^{12}$ L$_{\odot}$, while the
other Type 1 AGNs are less luminous. Models which can
explain the presence of the 10\,\mum\ silicate emission feature in
some type 1 AGNs will also need to explain their absence in others.

%%%%%%%%%%%%%%%%%%%%%%%%%%%%%%%%%%%%%%%%%%%%%%%%%%%%%%%%%%%%%%%%%%%%%

\acknowledgments

The authors wish to thank Bill Forrest, Elise Furlan and Terry Herter 
for their useful discussions.  Support for this work was provided by 
NASA through Contract Number 1257184 issued by the Jet Propulsion 
Laboratory, California Institute of Technology under NASA contract 
1407.  HWWS was supported under this contract through the 
Spitzer Space Telescope Fellowship Program.

%%%%%%%%%%%%%%%%%%%%%%%%%%%%%%%%%%%%%%%%%%%%%%%%%%%%%%%%%%%%%%%%%%%%%

\end{document}